\documentclass[a4paper,10pt]{article}

\usepackage{graphics,epsfig}
\usepackage{amsmath}
\usepackage{amsfonts}
\usepackage{amssymb}

\newcommand{\be}{\begin{equation}}
\newcommand{\ee}{\end{equation}}
\newcommand{\ba}{\begin{eqnarray}}
\newcommand{\ea}{\end{eqnarray}}
\def\tr{{\rm tr}}

\begin{document}

\title{Analysis of dilepton angular distributions in a parity breaking medium}
\author{A. A. Andrianov$^{1,2}$, V. A. Andrianov$^{1}$, D. Espriu$^{2}$ and X. Planells$^{2}$\\
\small{$^1$ V. A. Fock Department of Theoretical Physics,  Saint-Petersburg State University,}\\
{\small 198504 St. Petersburg, Russia}\\
{\small $^2$ Departament d'Estructura i Constituents de la Mat\`eria and }\\
{\small Institut de Ci\`encies del Cosmos (ICCUB),
Universitat de Barcelona,}\\
{\small  Mart\'\i \ i Franqu\`es 1, 08028 Barcelona, Spain}}

\date{}

\maketitle

\abstract{We investigate how local parity breaking due to large topological fluctuations may affect hadron physics. 
A modified dispersion relation is derived for the lightest vector mesons $\rho$ and $\omega$. They exhibit a 
mass splitting depending on their polarization. We present a detailed analysis of the angular distribution 
associated to the lepton pairs created from these mesons searching for polarization dependencies. We propose 
two angular variables that carry information related to the parity breaking effect. 
Possible signatures for experimental detection of local parity breaking that could potentially 
be seen by the PHENIX and STAR collaborations are discussed.

\vspace{-11cm}
\begin{flushright} ICCUB-14-042
\end{flushright}
\vspace{10.25cm}

\section{Introduction}
In recent years the behaviour of baryonic matter under extreme conditions has received a lot of attention \cite{Jacobs:2007dw,Blaizot:2011xf}. 
The medium created from heavy ion collisions (HIC) reaches thermodynamic regions that could not be explored before. 
In this context new properties of QCD are tested in current accelerator experiments such as RHIC and LHC \cite{Andronic:2009gj,Blaizot:2009ws}.

Some time ago effective Lagrangian studies for dense systems showed the possibility of spontaneous breaking of parity \cite{anesp}. A rigorous proof does not exist in QCD yet due to the difficulties of dealing with non-vanishing chemical potentials \cite{latt}. A different effect was proposed in \cite{kharzeev2,kharz}, where an isosinglet pseudoscalar condensate was shown to be likely to appear due to large topological fluctuations. Then, the presence of a non-trivial angular momentum (or magnetic field) in HIC could lead to the so-called Chiral Magnetic Effect (see the review of this and other related "environmental symmetry violations" in heavy ion collisions in \cite{Liao:2014ava}). It is also known that 
long-lived topological fluctuations can be treated in a quasi-equilibrium situation by means of an axial chemical potential $\mu_5$ \cite{kharz2}. Studies of simple models like Nambu--Jona-Lasinio or a generalised $\Sigma$-model have been addressed with a vector and an axial chemical potentials \cite{njl,epjc}.

In a recent work by the present authors \cite{plb} the main properties of the lightest vector mesons were described in the context of a local parity breaking (LPB) medium induced by $\mu_5\neq0$. It was shown that a distortion of the $\rho$ and $\omega$ spectral functions could be explained by a non-vanishing $\mu_5$\footnote{Distortion of the $\rho-\omega$ spectrum is well known to happen in HIC. Conventional explanations exist \cite{rapp,renkr,cassing,zahed} and if LPB occurs, all effects need to be treated jointly as long as they do not represent double counting.}. A remarkable polarization asymmetry appeared around their respective resonance poles. In this work we want to address a careful treatment of the polarization distribution associated to the lepton pairs produced in the decay of these mesons and to give possible signatures for experimental detection of LPB.

Current HIC experiments have investigated possible polarization dependencies in dilepton angular distributions with no significant results \cite{na60pol}. We claim in this work that the conventional angular variables used in such studies are not efficient enough to convey all the information related to the parity breaking effect. The dilepton invariant mass plays a key role in this game, which is normally missed. If not considered, dilepton polarization dependencies wipe out and no net effect can be extracted from experimental data. We will see that a combined analysis of some characteristic angles together with the dilepton invariant mass constitutes the appropriate framework to investigate the possible polarization asymmetry predicted by LPB.

This work is organised as follows: in Section 2, we introduce the Vector Meson Dominance model approach to LPB and derive the distorted dispersion relation that affects the lightest meson states $\rho$ and $\omega$. In Section 3 two different analysis of the dilepton angular distribution are presented in order to reflect how the LPB effect may be experimentally detected. Our conclusions are summarized in Section 4.

\section{Vector Meson Dominance approach with LPB}
The appropriate description of electromagnetic interactions of hadrons at low energies is the Vector Meson Dominance model containing the lightest vector mesons $\rho^0$ and $\omega$ in the $SU(2)$ flavour sector \cite{vmd,vmd2}. Quark-meson interactions are governed by
\begin{align}
\mathcal L_{\text{int}}=\bar q\gamma_\mu V^\mu q, \qquad V_\mu\equiv -eA_\mu Q+\frac12 g_\omega\omega_\mu \textbf{I}+\frac12 g_\rho\rho_\mu\tau^3,
\end{align}
where $Q=\frac12\tau^3+\frac16\textbf{I}$, $g_\omega\simeq g_\rho\equiv g\approx 6$. The kinetic and mass terms are given by
\begin{align}
\nonumber \mathcal L_{\text{kin}}&=-\frac14\left (F_{\mu\nu}F^{\mu\nu}+\omega_{\mu\nu}\omega^{\mu\nu} +\rho_{\mu\nu}\rho^{\mu\nu}\right )+\frac12 V_{\mu,a}m^2_{ab}V^\mu_b,\\
m^2_{ab}&=m_V^2\begin{pmatrix}
\frac{10e^2}{9g^2} & -\frac e{3g} & -\frac eg\\
-\frac e{3g} & 1 & 0\\
-\frac eg & 0 & 1
\end{pmatrix}, \qquad \det(m^2)=0,
\end{align}
where $V_{\mu,a}\equiv(A_\mu,\omega_\mu,\rho^0_\mu\equiv\rho_\mu)$ and $m_V^2=m_\rho^2= 2g_\rho^2f_\pi^2\simeq m_\omega^2$. The parity breaking effect is transmitted to the vector sector through the Chern-Simons (CS) term
\begin{align}
\mathcal L_{\text{CS}}=-\frac14 \varepsilon^{\mu\nu\rho\sigma}\tr(\hat\zeta_\mu V_\nu V_{\rho\sigma})=\frac12\tr(\hat\zeta\varepsilon_{jkl} V_j\partial_k V_l)=\frac12\zeta\varepsilon_{jkl}V_{j,a}N_{ab}\partial_k V_{l,b},
\end{align}
where a pseudoscalar time-dependent background is considered in order to accommodate an axial chemical potential associated to large topological fluctuations (for a detailed derivation see \cite{plb}). This development provides the relation $\zeta=N_cg^ 2\mu_5/8\pi^2\simeq 1.5\mu_5$. We will consider an isosinglet pseudoscalar background $\hat\zeta=2\zeta/g^2 \textbf I$ whose mixing matrix reads
\begin{equation}
N_{ab}=\begin{pmatrix}
\frac{10e^2}{9g^2} & -\frac e{3g} & -\frac eg\\
-\frac e{3g} & 1 & 0\\
-\frac eg & 0 & 1
\end{pmatrix}=\frac{m^2_{ab}}{m_V^2}.
\end{equation}
A non-vanishing axial chemical potential should be relevant in experiments where the nuclear fireball is rather hot but not very dense $T\gg\mu$, which is the expected scenario in HIC at LHC or RHIC.

After diagonalizing the system a distorted dispersion relation for vector mesons is found depending on their 3-momentum and polarization $\epsilon$
\begin{equation}\label{disp.rel}
m_{V,\epsilon}^2=m_V^2-\epsilon\zeta|\vec k|
\end{equation}
whereas photons remain undistorted. Transverse vector mesons ($\epsilon=\pm 1$) exhibit a shifted effective mass whereas longitudinal ones ($\epsilon=0$) are not affected by parity breaking. The splitting $m_{V,+}^2<m_{V,L}^2<m_{V,-}^2$ is a clear signature of LPB as well as Lorentz symmetry breaking due to the choice of a time-dependent background. The poles associated to $\pm$ polarized mesons depend on $|\vec k|$ and thus they appear as broadened resonances. This effect implies a reduction of the dilepton production at the nominal vacuum peak and an enhancement aside it related to the transverse polarized resonance peaks.

\section{Dilepton polarization analysis in $V\to\ell^+\ell^-$ decays}
The treatment of $\zeta$ in the dispersion relation shown in Eq. \eqref{disp.rel} is a non-trivial task. After a HIC, the system quickly goes to the QGP with a very short thermalization time (attempts to explain this process in holographic QCD can be found in \cite{holo}). LPB is expected to take place in the 
longer period when the hadronic phase is reentered. It is not clear that the value of $\zeta$ is uniform but we shall assume so here and consider an average $\zeta$ which has to be understood as the effective value extracted from such dynamics. In this work we will only consider $|\zeta|$ since a change of sign simply interchanges the $\pm$ polarizations but does not affect the separation of the polarization-dependent vector meson masses.

The dilepton production from the $V\to \ell^+\ell^-$ decays is governed by
\begin{align}\label{genericproduct}
\nonumber \frac{dN_V}{dM}=&c_V\frac{\alpha^2}{24\pi^2M}\left (1-\frac{n_V^2 m_\pi^2}{M^2}\right )^{3/2}\int\frac{d^3\vec k}{E_k}\frac{d^3\vec p}{E_p}\frac{d^3\vec p'}{E_{p'}}\delta^4(p+p'-k)\\
&\sum_{\epsilon}\frac{m_{V,\epsilon}^4\left (1+\frac{\Gamma_V^2}{m_V^2}\right )}{\left (M^2-m_{V,\epsilon}^2\right )^2+m_{V,\epsilon}^4\frac{\Gamma_V^2}{m_V^2}}  P^{\mu\nu}_{\epsilon}(M^2g_{\mu\nu}+4p_\mu p_\nu)\frac1{e^{M_T/T}-1},
\end{align}
where $V=\rho,\omega$ and $n_V=2,0$ respectively\footnote{$n_\omega=0$ is taken since we do not include the threshold to 3 pions.}, and $M>n_Vm_\pi$. $M_T$ is the transverse mass $M_T^2=M^2+\vec k_T^2$ where $\vec k_T$ is the vector meson transverse momentum and $M$ is the dilepton invariant mass. The projectors $P_\epsilon^{\mu\nu}$ are detailed in \cite{project}. $p$ and $p'$ correspond to the lepton and anti-lepton momenta. A Boltzmann distribution is included with an effective temperature $T$ \cite{lmlt,phenix}. A dilepton invariant mass smearing is taken into account in our computations but omitted in the previous formula and the following ones for the sake of simplicity. Finally, the constants $c_V$ normalize the contribution of the respective resonances (see \cite{plb}).

The temperature $T$ appearing in the previous formula is not the 'true' temperature of the hadron gas; it rather corresponds to the 
effective temperature $T_{flow}$ of the hadrons that best describes the slope of the multipliplicity distribution $d^3n/dp^3(M_T)$ for a given
range of invariant masses. Obviously
$T_{flow} > T_{true}$ and therefore it is quite possible that $T_{flow}$ exceeds the deconfinement temperature while one is still
dealing with hadrons. 

We will implement the experimental cuts of the PHENIX experiment ($p_T>200$ MeV and $|y_{ee}|<0.35$ in the center of mass frame
\footnote{We have not implemented the single electron cut $|y_{e}|<0.35$ because in practice it makes no visible difference with
simply imposing a cut on the dilepton pair momentum as a whole ---this has been checked explicitly}) and 
take $T\simeq 220$ MeV (see the previous comment) together with a gaussian invariant mass smearing of width 10 MeV \cite{phenix,masssmear} 
so as to investigate the angular distribution of the electron pairs produced from the meson decay. From now on the overall constants $c_V$ 
are chosen in such a way that after integrating the entire phase space the total production at the vacuum resonance peak is 
normalized to 1 unless otherwise is stated. This choice will help us to quantify the number of events found when the phase 
space is restricted in the following sections. We define the normalized number of events as
\begin{equation}
N_{\theta}(M)=\frac{\int_{\Delta\theta}\frac{dN}{dMd\cos\theta}\left (M,\cos\theta\right )d\cos\theta}{\int_{-1}^{+1}\frac{dN}{dMd\cos\theta}\left (M=m_V,\cos\theta\right )d\cos\theta},
\end{equation}
being $\theta$ one of the angles that we will investigate in the next sections.

Current angular distribution analysis \cite{Bratk,CS} are not the most suitable way to extract all the information related to the parity breaking effects we present. First, they omit the non-trivial dependence on $M$ where one could be able to see the resonances associated to the transverse polarized mesons. And second, the usual angular variables considered are not able to isolate the different polarizations. We will perform a two-dimensional study of the decay product with the dilepton invariant mass $M$ and a certain angle that we will introduce below. Our aim is to elucidate which other angular variables of the decay process should be taken into account when experimental data is analysed. Let us now consider two different angular variables.

\subsection{Case A}
The first variable we will investigate is the angle $\theta_A$ between the two outgoing leptons in the laboratory frame. Some basic algebraic manipulations lead us from Eq. \eqref{genericproduct} to
\begin{align}
\nonumber \frac{dN_V}{dMd\cos\theta_A}=&c_V\frac{\alpha^2}{6\pi M}\left (1-\frac{n_V^2m_\pi^2}{M^2}\right )^{3/2}\int\frac{p^2p'^2dpd\cos\theta d\phi}{E_p\sqrt{(M^2-2m_\ell^2)^2-4m_\ell^2(E_p^2-p^2\cos^2\theta_A)}}\\
&\sum_{\epsilon}\frac{m_{V,\epsilon}^4\left (1+\frac{\Gamma_V^2}{m_V^2}\right )}{\left (M^2-m_{V,\epsilon}^2\right )^2+m_{V,\epsilon}^4\frac{\Gamma_V^2}{m_V^2}}  P^{\mu\nu}_{\epsilon}(M^2g_{\mu\nu}+4p_\mu p_\nu)\frac1{e^{M_T/T}-1},
\end{align}
where $\theta$ is the angle between $p$ and the beam axis (not to be confused with $\theta_A$ previously defined) and $\phi$ is the azimuthal angle of $p'$ with respect to $p$. The lepton energies are given by $E_{p}^2-p^2=E_{p'}^2-p'^2=m_\ell^2$ and $p'=|\vec p'|(\theta_A,p,M,m_\ell)$ can be found from the decay kinematics that lead to the following equation:
\begin{equation}
E_pE_{p'}-pp'\cos\theta_A=\frac{M^2}2-m_\ell^2.
\end{equation}

In the left panels of Fig. \ref{a} we present the results of the dilepton production in the $\rho$ channel as a function of its invariant mass $M$ integrating small bins of $\Delta\cos\theta_A=0.2$. The different curves displayed in the plot correspond to $\cos\theta_A\in [-0.2,0],[0,0.2],[0.2,0.4],[0.4,0.6]$ and $[0.6,0.8]$. The upper panel corresponds to the vacuum case with $\mu_5=0$ while the lower one represents a parity-breaking medium with $\mu_5=300$ MeV (see \cite{njl,epjc} for a justification of such scale). The LPB effect produces a secondary peak corresponding to a transverse polarization in the low invariant mass region in addition to the vacuum resonance. If the effective temperature $T$ increased, we would see an enhancement of the upper mass tail making the secondary peak smaller in comparison to the vacuum one. As $\cos\theta_A$ increases, the outgoing leptons are more collimated, meaning a higher meson 3-momentum and from the dispersion relation in Eq. \eqref{disp.rel}, a larger separation between the two peaks together with a Boltzmann suppression. Note that the highest peaks in the figures with $\mu_5\neq0$ have a value $\simeq 0.1$, this is a $10\%$ of the total production at the vacuum peak when the total phase space is considered (except for the detector cuts). We took a representative value $\mu_5=300$ MeV so as to show two different visible peaks. If the axial chemical potential acquires smaller values, the secondary peak approaches the vacuum one. The smaller $\mu_5$ is taken, the more collinear the leptons have to be in order to obtain a visible and significant secondary peak (recall Eq. \eqref{disp.rel}).

\begin{figure}[h!]
\centering
\includegraphics[scale=0.35]{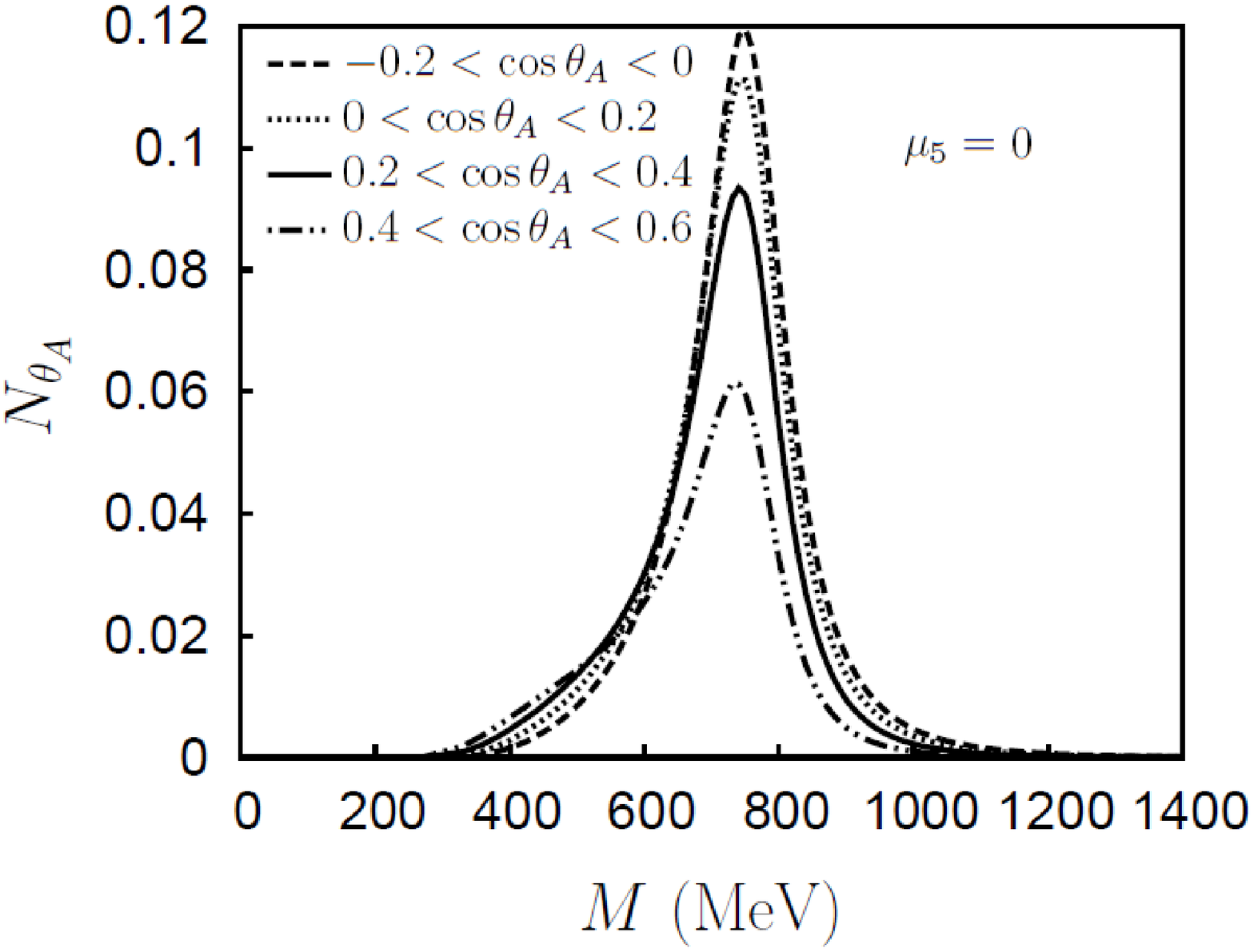}\quad \includegraphics[scale=0.35]{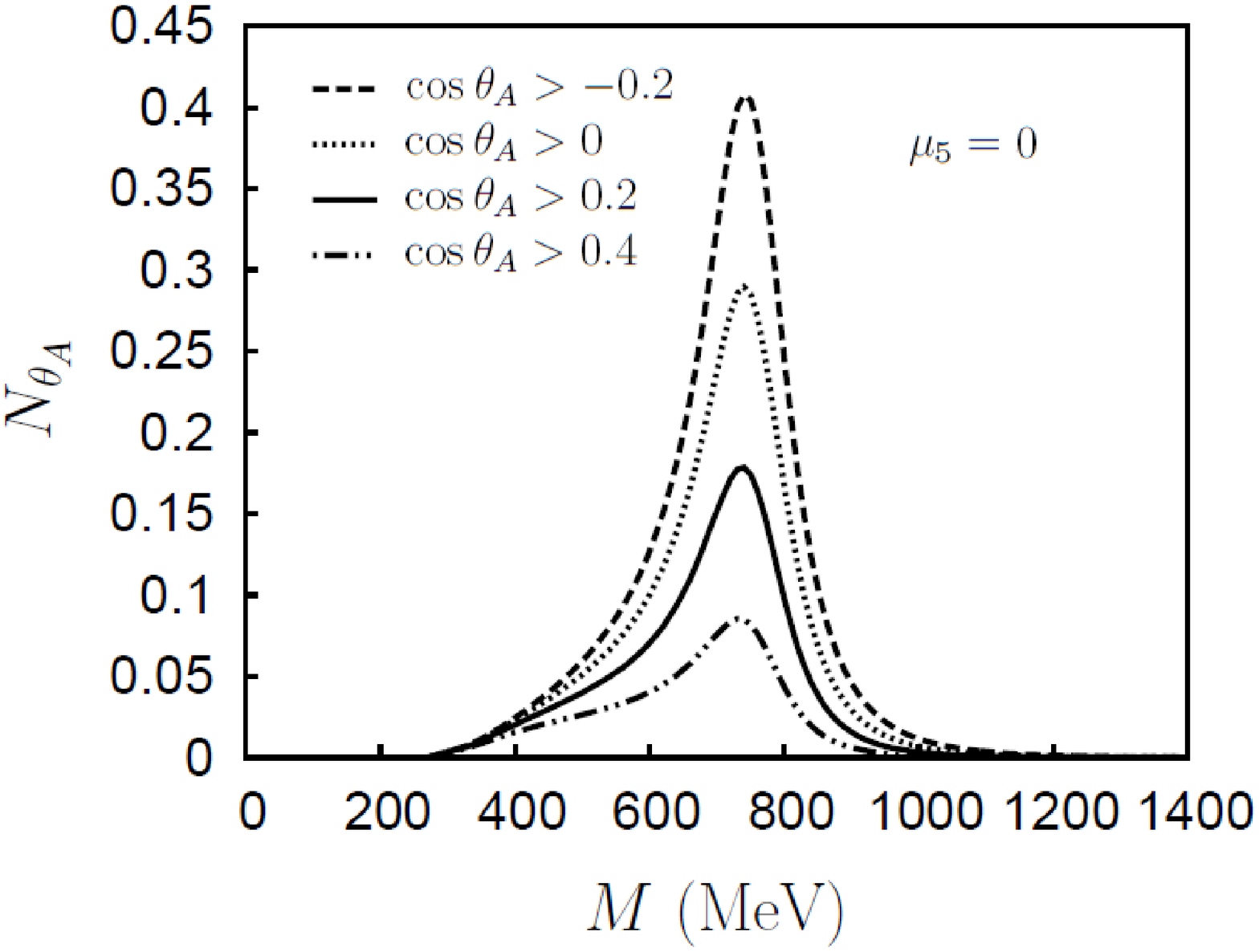}
\includegraphics[scale=0.35]{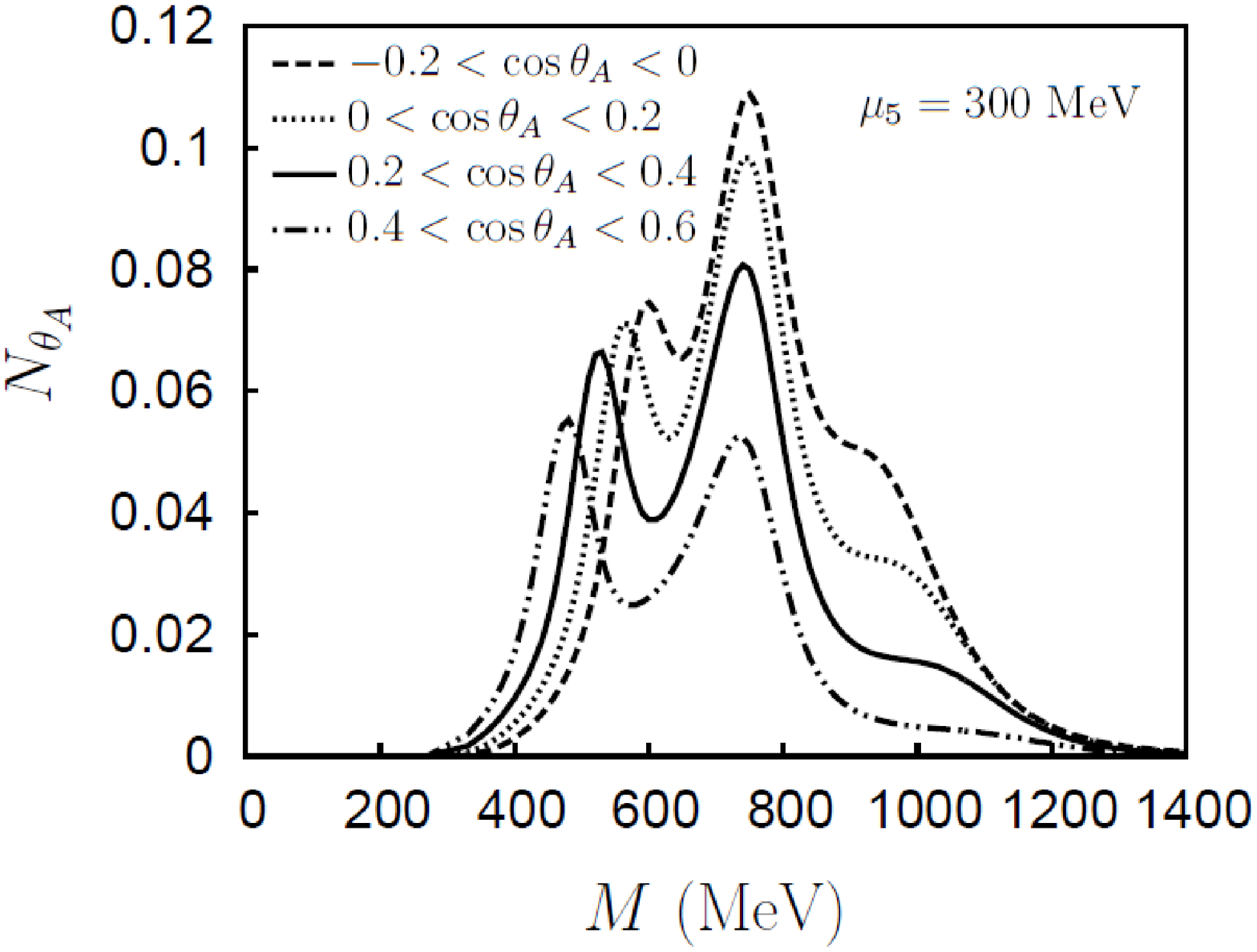}\quad \includegraphics[scale=0.35]{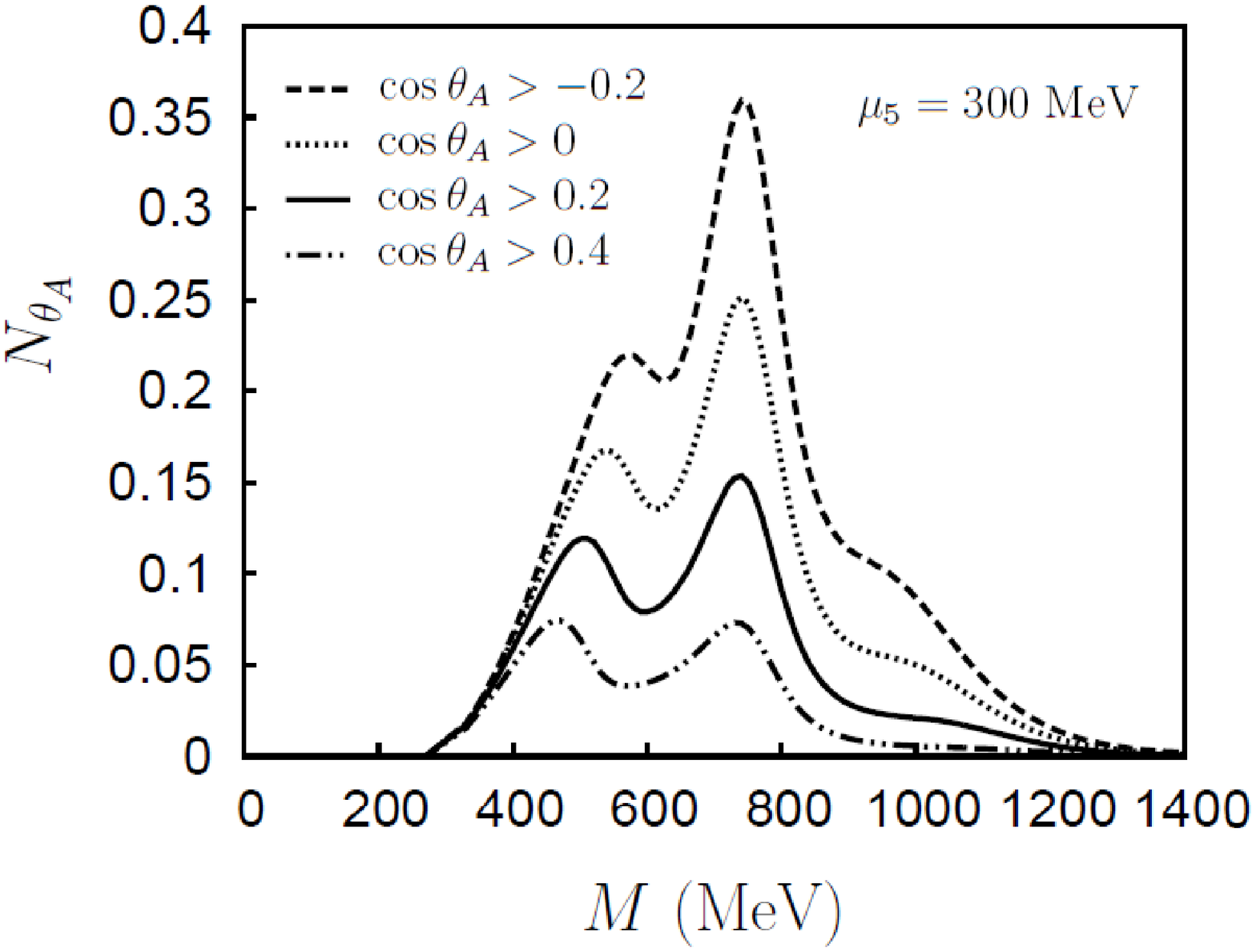}
\caption{The $\rho$ spectral function is presented depending on the invariant mass $M$ in vacuum ($\mu_5=0$) and in a parity-breaking medium with $\mu_5=300$ MeV (upper and lower panels, respectively) for different ranges of the angle between the two outgoing leptons in the laboratory frame $\theta_A$. We display the curves corresponding to $\cos\theta_A\in [-0.2,0],[0,0.2],[0.2,0.4],[0.4,0.6]$ and $[0.6,0.8]$ in the left panels, and $\cos\theta_A\geq -0.2,0,0.2,0.4$ in the right ones. The total production at the vacuum peak is normalized to 1 when the entire phase space is considered. Results are presented for the experimental cuts quoted by PHENIX \cite{phenix}.}\label{a}
\end{figure}

Dealing with quantities smaller than $10\%$ may be tricky. As the main information related to LPB is focused in $\cos\theta_A\approx 1$, in the right panels of Fig. \ref{a} we show a somewhat more experimentally oriented set of plots where we integrated $\cos\theta_A$ down to -0.2, 0, 0.2 and 0.4. The upper panel corresponds to vacuum and the lower one represents a parity-breaking medium with $\mu_5=300$ MeV. The number of events grows when more and more separated leptons are considered but the secondary peak smears and becomes much less significant. Therefore, an optimization process should be performed in every different experiment so as to find the most suitable angular coverage.

Another issue that could be experimentally addressed is the analysis of $\mu_5$ for a particular (and fixed) coverage of $\theta_A$. If the secondary peaks due to the transversally polarized mesons were found, their positions would be an unambiguous measurement of $\mu_5$ (more precisely, $|\mu_5|$). In Fig. \ref{arw} we integrate the forward direction of the two outgoing leptons, i.e. $\cos\theta_A\geq 0$, and examine how the transverse polarized peaks move with respect to the vacuum one when we change the values of the axial chemical potential. The $\rho$ and $\omega$ spectral functions are displayed in the upper panels for $\mu_5=100,200$ and 300 MeV. The same tendency is found in both graphics except for the fact that both transverse peaks are observed in the $\omega$ channel (one below and one above the vacuum resonance). For small values like $\mu_5\simeq 100$ MeV the vacuum peak hides the transverse one due to the $\rho$ width, being impossible to discern if one or two resonances are present. In the $\omega$ channel all the peaks are visible even for such small $\mu_5$. The latter channel could be the most appropriate one in order to search for polarization asymmetry. For completeness, we also present their combination in the lower panel normalized to PHENIX data. The normalized number of events are defined in this case as
\begin{equation}
N_{\theta}^{\text{PH}}(M)=\frac{dN^{\text{PH}}}{dM}\left (M=m_V\right )\frac{\int_{\Delta\theta_A}\frac{dN}{dMd\cos\theta}\left (M,\cos\theta\right )d\cos\theta}{\int_{-1}^{+1}\frac{dN}{dMd\cos\theta}\left (M=m_V,\cos\theta\right )d\cos\theta},
\end{equation}
where $\frac{dN^{\text{PH}}}{dM}(M)$ is the theoretical spectral function used in the PHENIX hadronic cocktail. An enhancement factor 1.8 is 
included for the $\rho$ channel due to its regeneration into pions within the HIC fireball (a plausible consequence 
of the "$\rho$ clock" effect \cite{Heinz:1991fn,Specht:2010xu}). 
There are no published reports of a direct determination of the $\rho/\omega$ ratio at PHENIX and we
have decided to use the above value that is very close to the average value for the ratio of the respective cross sections reported
by NA60 \cite{scomparin,arnaldi2} and was used by us in previous works after a fit to the hadronic cocktail. 
We note that the $\eta/\omega$ ratios measured both by NA60 and PHENIX collaborations roughly agree. 
Note too that due to the assumed effective thermal distribution and the fact that the $\rho$
and $\omega$ are nearly degenerate in mass the ratio  $c_\rho/c_\omega$ is identical to the ratio of the respective cross-sections.  
However it should be stated right away that our conclusions do not depend substantially on the precise value of the ratio  $c_\rho/c_\omega$ as the
experimental signal that we propose is amply dominated by the $\omega$ when the two resonances are considered together (see. Fig. \ref{arw}).

\begin{figure}[h!]
\centering
\includegraphics[scale=0.35]{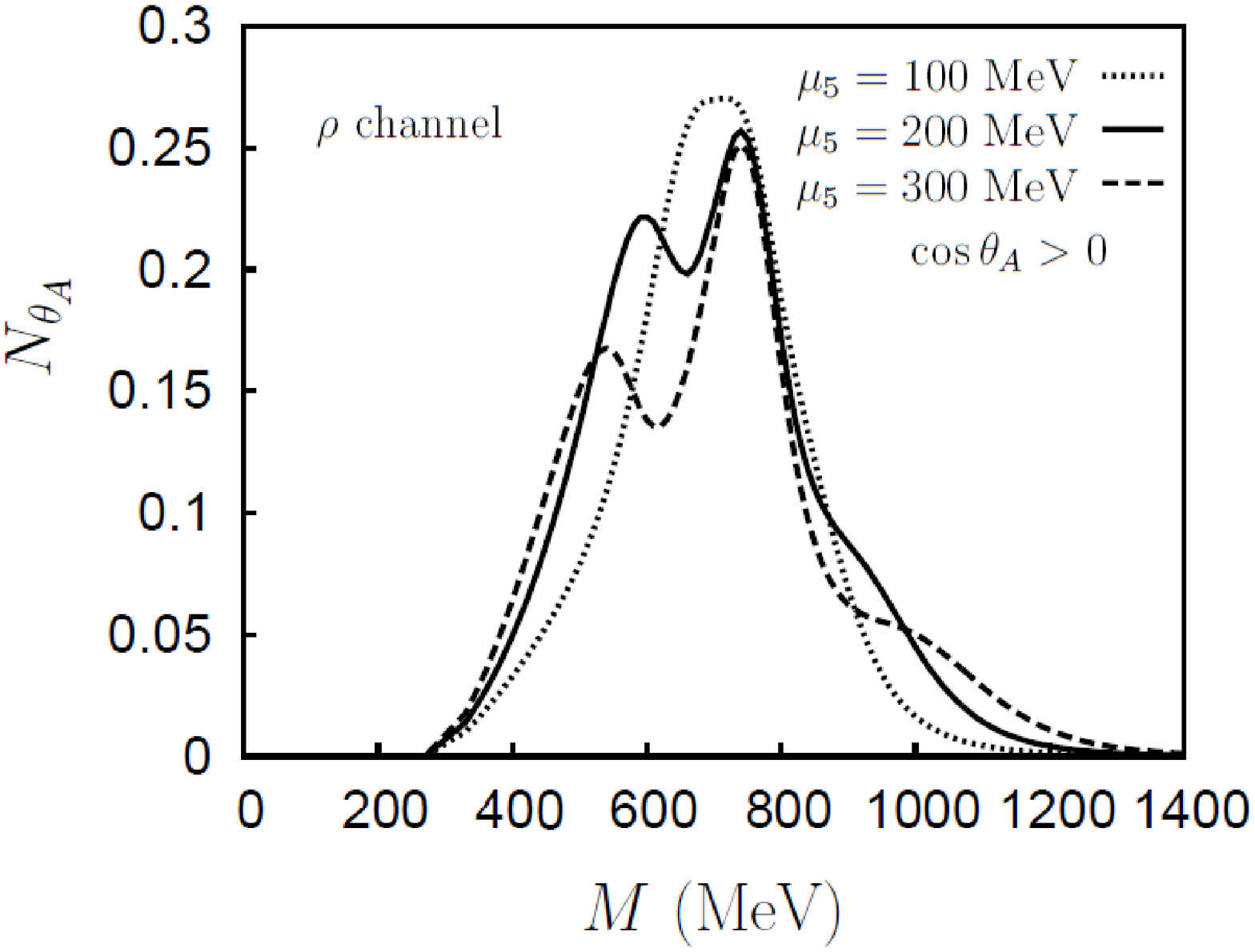}\qquad \includegraphics[scale=0.35]{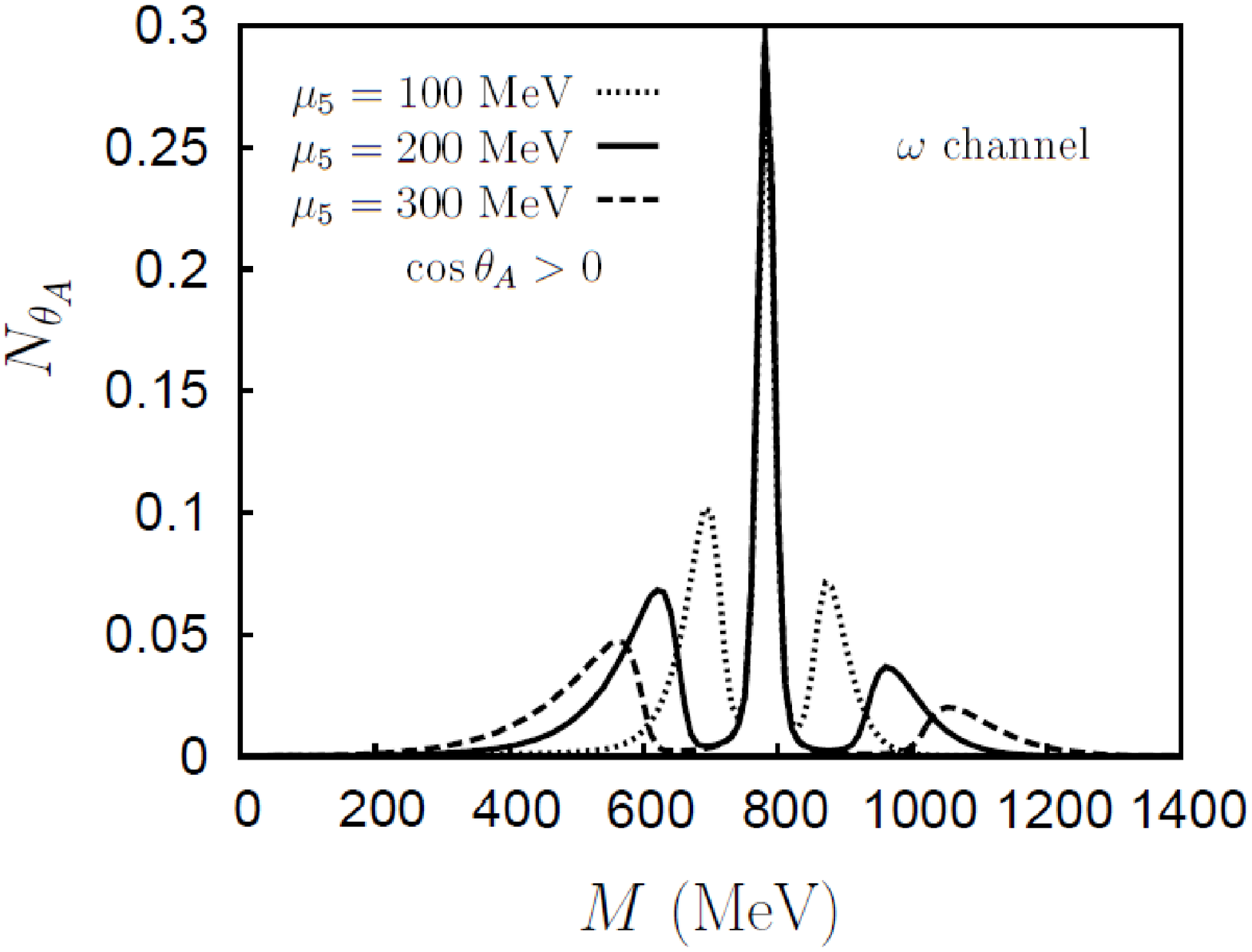}\\
\includegraphics[scale=0.45]{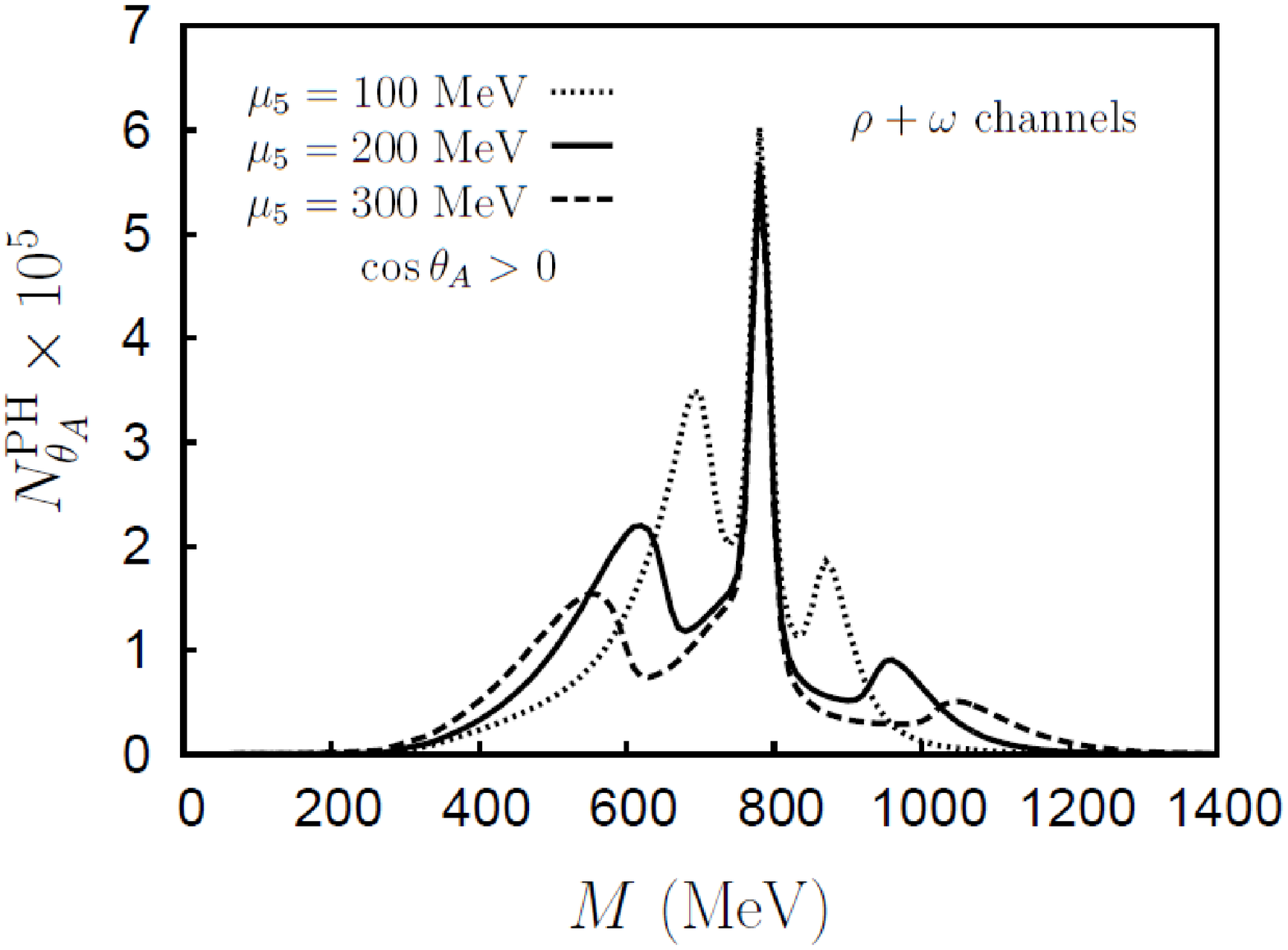}
\caption{The $\rho$ (upper-left panel) and $\omega$ (upper-right panel) spectral functions and their combination (lower panel) are showed depending on the invariant mass $M$ and integrating the forward direction $\cos\theta_A\geq 0$ for $\mu_5=100,200$ and 300 MeV. In the upper panels the total production at the vacuum peak is normalized to 1 when the entire phase space is considered whereas the lower panel is normalized to PHENIX data. Results are presented for the experimental cuts quoted by PHENIX \cite{phenix}.}\label{arw}
\end{figure}

\subsection{Case B}
The second variable we present is the angle $\theta_B$ between one of the two outgoing leptons in the laboratory frame and the same lepton in the dilepton rest frame.  The dilepton production is given by
\begin{align}
\nonumber \frac{dN_V}{dM}=&c_V\frac{\alpha^2}{6\pi M}\left (1-\frac{4m_\pi^2}{M^2}\right )^{3/2}\int\frac{k^2dkd\cos\theta}{E_k}\frac{p^2d\cos\psi d\phi}{\sqrt{M^4-4m_\ell^2(E_k^2-k^2\cos\psi^2)}}\\
&\sum_{\epsilon}\frac{m_{V,\epsilon}^4\left (1+\frac{\Gamma_V^2}{m_V^2}\right )}{\left (M^2-m_{V,\epsilon}^2\right )^2+m_{V,\epsilon}^4\frac{\Gamma_V^2}{m_V^2}} P^{\mu\nu}_{\epsilon}  (M^2g_{\mu\nu}+4p_\mu p_\nu)\frac1{e^{M_T/T}-1},
\end{align}
where $k(p)$ and $E_k(E_p)$ are the meson (selected lepton) momentum and energy. $\psi$ and $\phi$ are the polar and azimuthal angles respectively of the lepton with respect to the meson momentum. The beam direction forms an angle $\theta$ with the vector meson. One may extract $p$ from the decay kinematics. All these variables are defined in the laboratory frame. The new angle $\theta_B$ is defined via
\begin{equation}
\frac M2\cos\theta_B\sqrt{M^2-4m_\ell^2}=p(M\sin^2\psi+E_k\cos^2\psi)-kE_p\cos\psi.
\end{equation}
Regarding the fact that we would like to perform experimental cuts in $\theta_B$, our numerical computations integrate the whole phase space of the decay and reject the regions with unwanted values of $\theta_B$ instead of treating this angle as an integration variable, which makes the calculations more complicated.

Events with $\cos\theta_B\simeq 1$ correspond to one lepton being produced in approximately the same direction in the laboratory frame and in the meson rest frame, implying that in the laboratory frame the vector meson is almost at rest. Vector mesons with low momentum are not suppressed by the Boltzmann weight but do not carry relevant information about $\mu_5$. Therefore the opposite limit (high momenta) will be the important one for our purposes.

In the left panel of Fig. \ref{beta} the $\rho$ spectral function is presented in small bins of $\Delta\cos\theta_B=0.1$. The curves correspond to $\cos\theta_B\in[0.3,0.4], [0.4,0.5], [0.5,0.6]$ and $[0.6,0.7]$ with a fixed $\mu_5=300$ MeV. At first glance, the plot looks as the one showed in the previous section for $\theta_A$ with a similar secondary peak below the vacuum resonance. The main difference with the previous section is the number of events. The analysis of $\theta_B$ is more sensitive to the Boltzmann suppression than the previous case with $\theta_A$. In the right panel of Fig. \ref{beta} it may be readily checked that the number of events at the highest secondary peak is $0.14$, whereas in Fig. \ref{a} it corresponds to $0.22$, a considerable enhancement of around a $60\%$.

\begin{figure}[h!]
\centering
\includegraphics[scale=0.35]{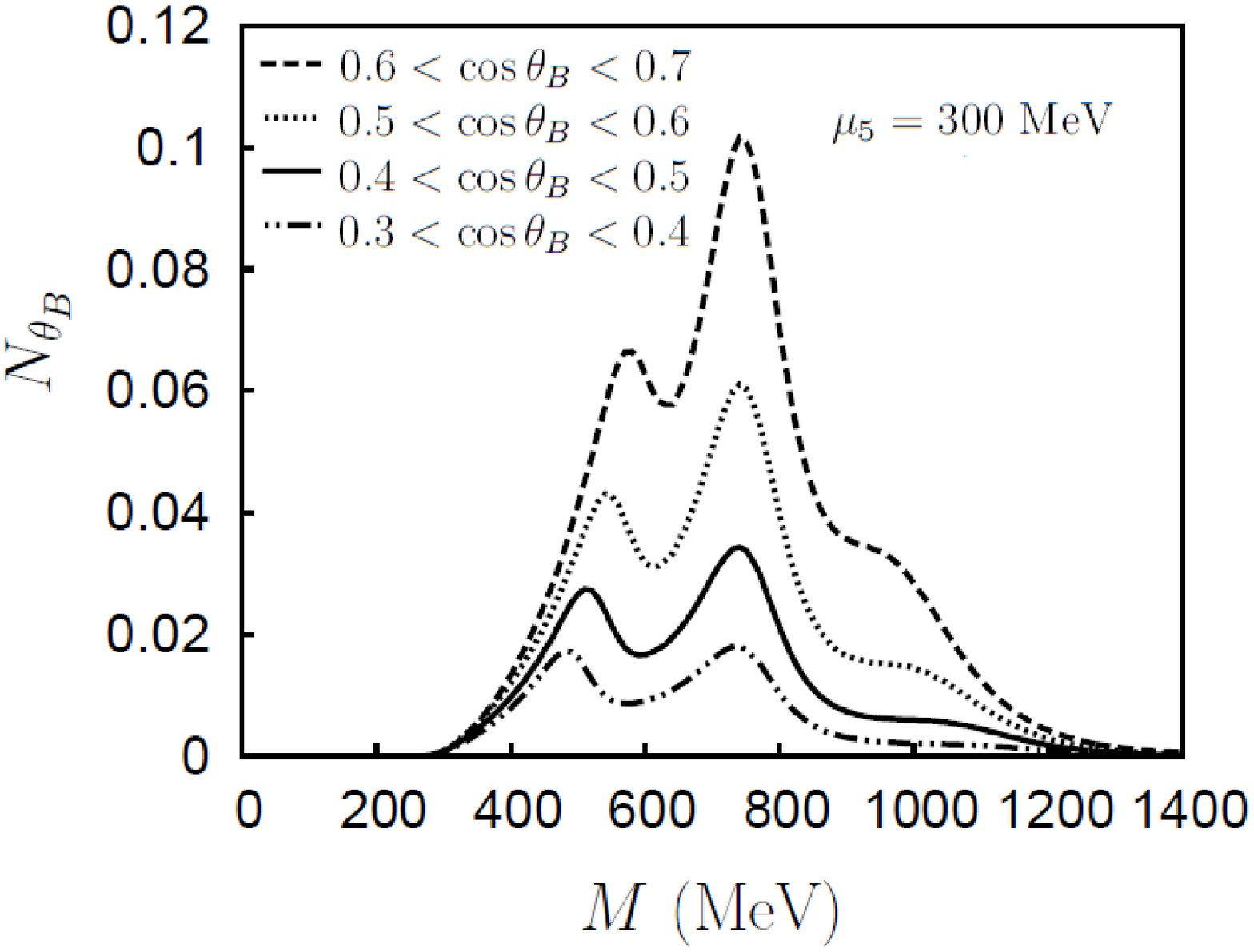}\quad \includegraphics[scale=0.35]{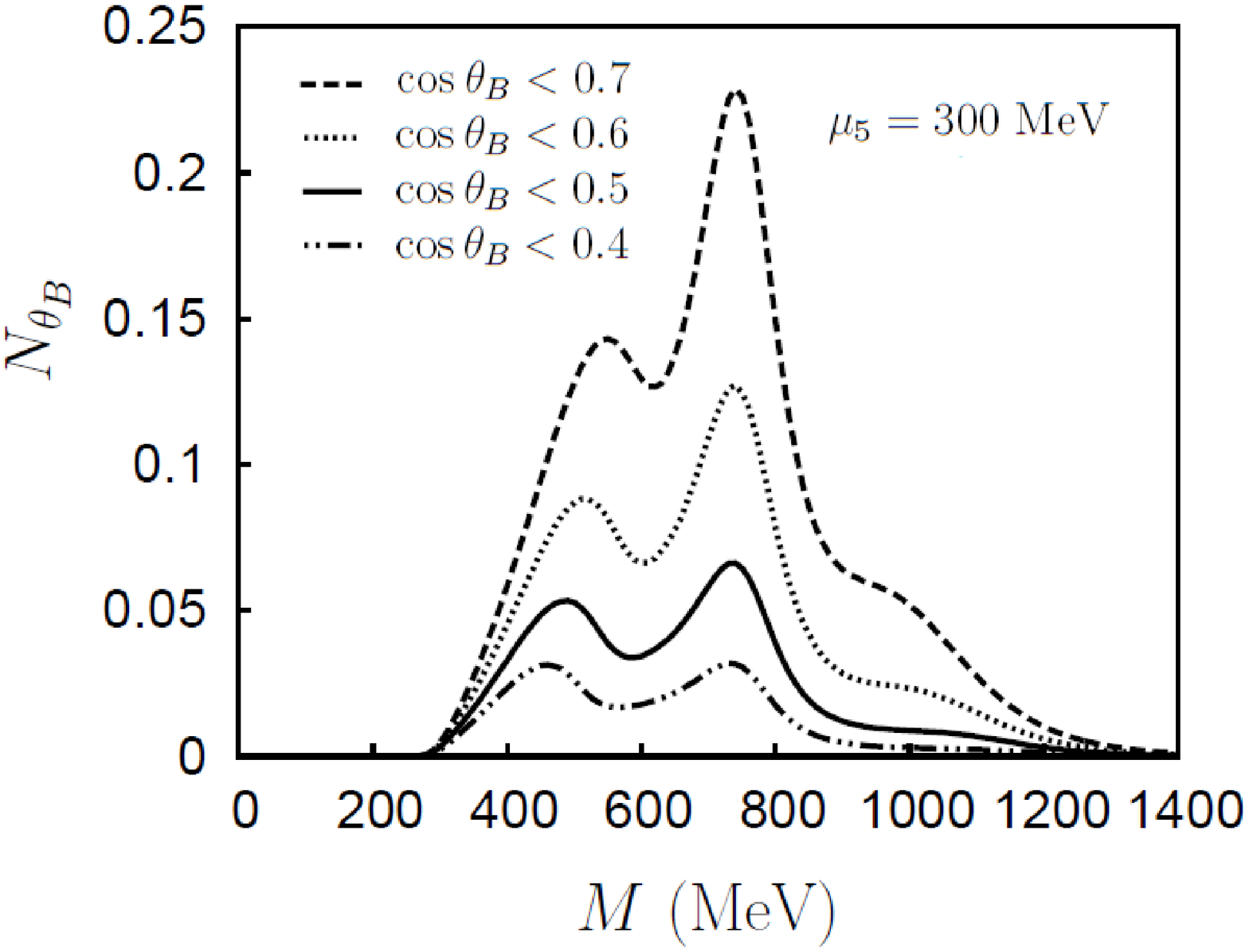}
\caption{The $\rho$ spectral function is presented depending on the invariant mass $M$ for different ranges of the angle $\theta_B$ between one of the outgoing leptons in the laboratory frame and the same lepton in the dilepton rest frame for fixed $\mu_5=300$ MeV. We display the curves corresponding to $\cos\theta_B\in [0.3,0.4],[0.4,0.5],[0.5,0.6]$ and $[0.6,0.7]$ in the left panel, and $\cos\theta_B\leq 0.4,0.5,0.6,0.7$ in the right one. The total production at the vacuum peak is normalized to 1 when the entire phase space is considered. Results are presented for the experimental cuts quoted by PHENIX \cite{phenix}.}\label{beta}
\end{figure}

A phenomenological analysis of $\mu_5$ depending on the position of the secondary peak as well as a comparison with the vacuum contributions may be equally described in this section but no new features are found so we omit the corresponding details.

\section{Conclusions}
In the context of a dense medium created from a heavy ion collision parity may be locally broken due to large topological fluctuations. Vector mesons acquire an effective mass depending on their polarization, 3-momentum and the axial chemical potential. Such a distorted dispersion relation predicts that massive vector mesons split into three polarizations with different masses. The resonance poles associated to the transverse mesons appear separated of vacuum peak implying a polarization asymmetry.

As possible signatures for experimental detection of LPB we presented a description in two angular variables that are not considered in 
the literature when angular distribution analysis are investigated. The first one is the angle between 
the two outgoing leptons produced from the meson decay in the laboratory frame. The second one is defined as the angle between 
one of the leptons at the laboratory frame and the same lepton in the dilepton rest frame. These angles are the most 
suitable ones so as to extract information about LPB. Without a careful choice of angular observables the effects of LPB on the
different polarizations can be easily missed.

We claim that a two-dimensional study of the decay product with the variables angle-dilepton mass could allow to distinguish 
at least two of the resonance poles and confirm or disprove the parity breaking hypothesis. We displayed the $\rho$ and $\omega$ 
spectral functions in order to illustrate this effect and discuss the most efficient ways to search for polarization asymmetry.

The study presented here is still not fully realistic and several improvements could be made such as e.g. including 
a complete hydrodynamical treatment, finite volume effects, thermal broadening of the resonances and a more detailed study of the
time and (effective) temperature dependence. Yet none of these are expected to erase the traces of LPB if the latter is present
and we would like to encourage the experimental collaborations to actively search for this interesting possibility.

\section*{Acknowledgements}
We acknowledge the financial support from projects FPA2010-20807, 2009SGR502, CPAN (Consolider CSD2007-00042).
A. A. and V. A. Andrianov are also supported by Grant RFBR project 13-02-00127 as well
as by the Saint Petersburg State University grants 11.38.660.201, 11.42.1292.2014 and 11.42.1293.2014.
X. Planells acknowledges the support from Grant FPU AP2009-1855.

\end{document}